# Charge Transport along the *c*-Axis in High-$T_c$ Cuprates*

Yoichi Ando

Central Research Institute of Electric Power Industry, Komae, Tokyo 201-8511, Japan

**Abstract**: Using 61-T pulsed magnetic fields, the normal-state $\rho_{ab}$ and $\rho_c$ are measured in Bi-2201 system down to 0.66 K, and the coexistence of the "metallic" $\rho_{ab}$ and the "semiconducting" $\rho_c$, usually called the charge confinement behavior, was confirmed to extend far below $T_c$. Recent measurement of the *c*-axis magnetoresistance under 16 T dc magnetic field in heavily underdoped Y-123 crystals revealed that the peculiar *c*-axis charge transport, and thus the charge confinement, is fundamentally related to the antiferromagnetic spin fluctuations.

## INTRODUCTION

One of the most unusual normal-state properties of the high-$T_c$ cuprates is the temperature dependence of the anisotropic resistivity: in samples near optimum doping, the in-plane resistivity $\rho_{ab}$ decreases linearly with decreasing $T$ over a wide temperature range, while the out-of-plane resistivity $\rho_c$ increases rapidly at low temperatures. To elucidate whether this contrasting behavior extends far below $T_c$ and thus is truly a "ground state" property of the normal state in the absence of superconductivity, we measured the normal-state $\rho_{ab}$ and $\rho_c$ in La-doped $Bi_2Sr_2CuO_y$ (Bi-2201) crystals by suppressing superconductivity with 61-T pulsed magnetic fields [1]. In sufficiently clean samples, we confirmed that the metallic $\rho_{ab}$ coexists with a "semiconducting" $\rho_c$ down to the lowest experimental temperature, 0.66 K, giving evidence of the non-Fermi-liquid nature of the cuprates. This result also evidences the existence of a charge confinement mechanism that persists to low temperatures. To further investigate the charge confinement, we have recently measured the *c*-axis magnetoresistance in heavily underdoped $YBa_2Cu_3O_{6+x}$ (Y-123) crystals that show Néel order. It was found that the peculiar *c*-axis transport is strongly related to the antiferromagnetism, because the *c*-axis magnetoresistance shows a step-like increase across the Néel temperature [2]. Such results strongly suggest that the charge transport along the *c*-axis, which takes place as a hopping process across the charge-confining $CuO_2$ planes, is fundamentally related to the antiferromagnetic spin fluctuations.

## CHARGE CONFINEMENT IN THE ZERO TEMPERATURE LIMIT

The contrasting behavior of $\rho_{ab}$ and $\rho_c$ has been the subject of intense study, both theoretically and experimentally [3], and is often counted as an evidence for the non-Fermi-liquid nature of the cuprates [4]. However, there have been discussions that the contrasting behavior might be a finite temperature effect in a Fermi liquid, where, for example, *c*-axis transport is due to phonon-assisted hopping [5]. Clearly, such question must be settled by measurements of the normal-state resistivity at low temperatures.

The most straightforward way to measure the normal-state resistivity below $T_c$ is to suppress superconductivity with an intense magnetic field. We suppressed superconductivity in La-doped

Bi-2201 single crystals using a 61 T pulsed magnetic field and measured both $\rho_{ab}$ and $\rho_c$ in the normal-state down to 0.66 K. The Bi-2201 system we used is a particularly suitable compound for studying the contrasting behavior: $T_c$ is relatively low among the high-$T_c$ cuprates; $\rho_c$ shows a strong divergence [6]; and the linear-$T$ behavior in $\rho_{ab}$ is quite robust (extends up to 700 K [6]). The $Bi_2Sr_{2-x}La_xCuO_y$ single crystals we report here are slightly overdoped, with nominal $x$=0.05 and the midpoint $T_c$ of about 13 K. The crystals were grown by the flux method and the cleanest ones show the in-plane resistivity $\rho_{ab}$ of about 130 μΩcm at 100 K. To measure the anisotropic resistivity, we employ a six-terminal method: two current contacts are located on the top *ab* face of the crystal with a pair of voltage contacts placed in between; an additional pair of voltage contacts is placed on the bottom face directly beneath the top-face voltage pair. Analyzing the top- and bottom-face voltages using the linear anisotropic resistivity model of Busch *et al.* [7] gives $\rho_{ab}$ and $\rho_c$.

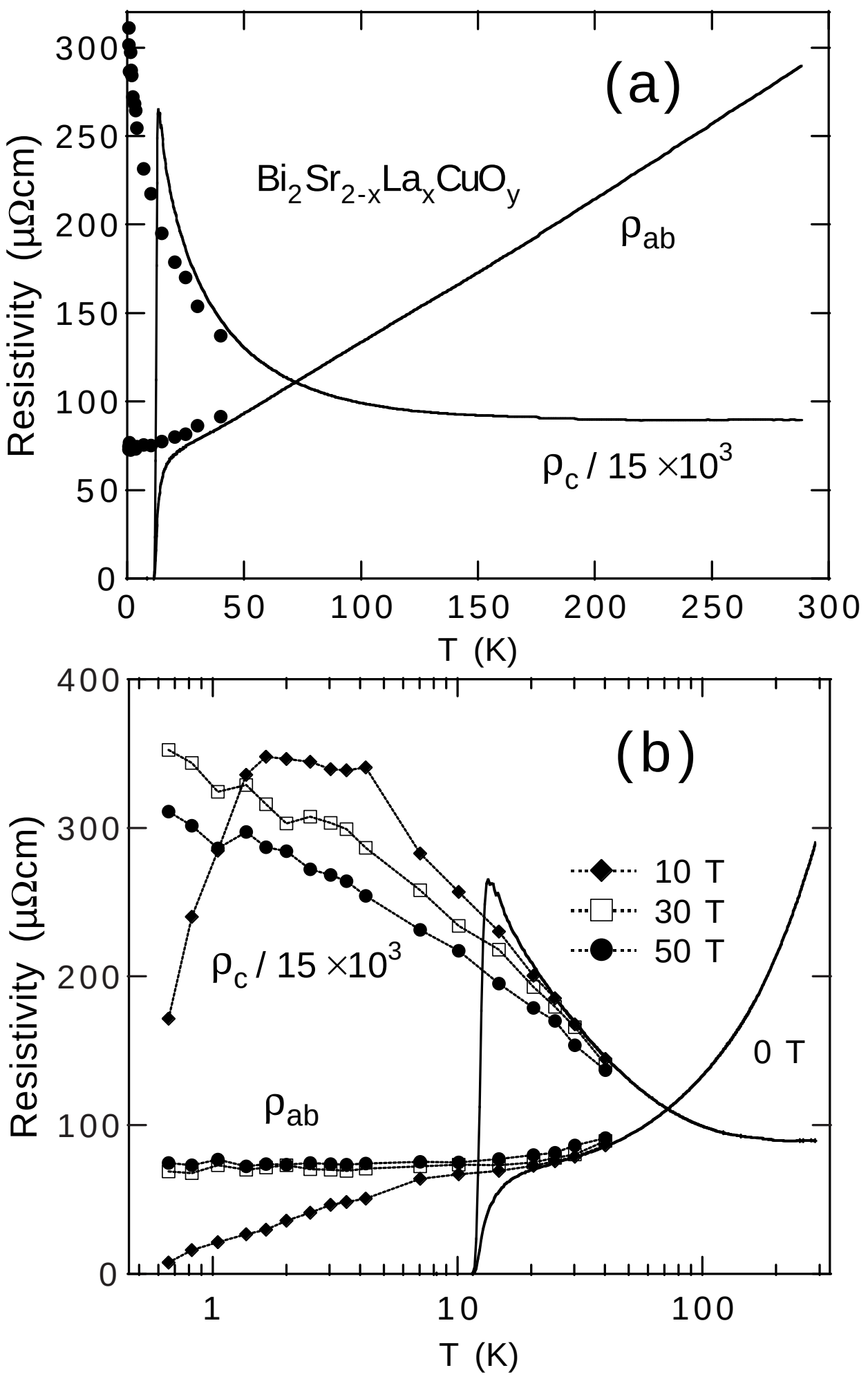

Fig. 1. (a) $T$ dependence of $\rho_{ab}$ and $\rho_c$ of one of the cleanest Bi-2201 samples in 0 and 60 T. (b) log$T$ plot of $\rho_{ab}$ and $\rho_c$ for fixed values of magnetic field, emphasizing the metallic $\rho_{ab}$ and diverging $\rho_c$ in the zero temperature limit. $\rho_c$ data are divided by $15\times10^3$.

The result of the high magnetic field measurement of one of the cleanest samples is shown in Fig. 1(a). In this sample, $\rho_{ab}$ stays metallic down to the lowest experimental temperature, 0.66 K, and $\rho_c$ continues to diverge. Therefore, the contrasting behavior of $\rho_{ab}$ and $\rho_c$ persists down to $T/T_c$ = 0.05. This strongly suggests that the metallic in-plane conduction and "semiconducting" out-of-plane conduction can indeed coexist in the zero-temperature limit when the in-plane disorder is sufficiently small [1]. Note that $\rho_{ab}$ in this Bi-2201 sample becomes as small as 74 μΩcm, which corresponds to $k_Fl \approx 42$ in the 2D model ($k_Fl = hc_0/\rho_{ab}e^2$, where $c_0$ = 12 Å is the interlayer distance). The data suggest that $\rho_{ab}$ is saturating to a residual resistivity, although we cannot exclude the possibility that $\rho_{ab}$ will cross over to an insulating behavior below our experimental temperature range. Ordinarily, however, such a large value of $k_Fl$ would assure metallic behavior to extremely low temperatures. Figure 1(b) shows a log$T$ plot of $\rho_{ab}$ and $\rho_c$ for various fixed magnetic fields in order to emphasize the low-temperature behavior. The large negative MR in $\rho_c$ is evident and $\rho_c$ diverges roughly logarithmically at low temperatures, comparable to the behavior reported in $La_{2-x}Sr_xCuO_4$ [8]. On the other hand, $\rho_{ab}$ below $T_c$ shows almost no temperature dependence and little magnetoresistance in this clean sample. This suggests that the *c*-axis transport is uncorrelated with the in-plane transport. This behavior of Bi-2201 contrasts strongly with that of $La_{2-x}Sr_xCuO_4$, where $\rho_{ab}$ becomes insulating whenever $\rho_c$ is diverging at low temperatures [8,9].

Let us discuss the implication of the above result. A clear indication is that the *c*-axis transport is incoherent even at very low temperatures. Within the framework of the Fermi-liquid theory, several models have been proposed [3] to explain the incoherence, including renormalization of the interlayer hopping rate [10], dynamical dephasing [11], and interlayer scattering [5,12]. Accounting for a "semiconducting" $\rho_c$ in these Fermi-liquid models in addition to the incoherence is more difficult and two possibilities have been discussed: phonon-assisted hopping [5,10] and temperature-dependent suppression of the density of states at the Fermi energy, $N(0)$ [13]. Rojo and Levin have found that phonon-assisted hopping can give a "semiconducting" $\rho_c$ over a broad temperature range, as long as the phonon energies are sufficiently large in helping electrons to hop across the planes. In this mechanism, however, $\rho_c$ must eventually become proportional to $\rho_{ab}$ as the characteristic phonon energy becomes small at low temperatures, roughly $T \leq 20$ K (an energy scale which is approximately 1/40 of the highest energy *c*-axis phonons [5]). In our data, $\rho_c$ continues to diverge at our lowest experimental temperature, 0.66 K, which clearly speaks against the phonon-assisted hopping model. The data pose an additional difficulty for Fermi-liquid models: the fact that $\rho_{ab}$ shows almost no temperature dependence and little magnetoresistance at low temperatures, while $\rho_c$ continues to show strong temperature and magnetic-field dependences. In the interlayer scattering models [5,12], the temperature dependence of $\rho_c$ is determined by the scattering time from the interplane (off-diagonal) disorder, $\tau_1$, and also by the scattering time from the in-plane (diagonal) impurities, $\tau_2$. Since the same scattering times enter into the expression of $\rho_{ab}$ in these interlayer scattering models, it is difficult to account for the temperature-independent $\rho_{ab}$. Alternatively, Zha, Cooper and Pines have proposed [13] that the low-temperature upturn in $\rho_c$ is due to a reduction in $N(0)$. Although there is no microscopic calculation of $\rho_{ab}$ in this model, one would expect $\rho_{ab}$ to be temperature dependent when $N(0)$ changes with temperature.

On the other hand, in non-Fermi-liquid theories of the high-$T_c$ cuprates, the incoherence of the *c*-axis conduction results from the in-plane quasiparticle confinement [4,14]. Both the resonating valence bond theory [15,16] and the Luttinger liquid theory [17] give metallic $\rho_{ab}$ accompanied by "semiconducting" $\rho_c$ in the zero-temperature limit. To summarize, the observation of coexistence of metallic $\rho_{ab}$ and "semiconducting" $\rho_c$ down to the lowest experimental temperature, 0.66 K, points toward a non-Fermi-liquid ground state in the high-$T_c$ cuprates.

## *c*-AXIS CHARGE TRANSPORT AND THE SPIN FLUCTUATIONS

A fundamental role of the magnetic interactions in the c-axis transport was recently found in a study of heavily underdoped $RBa_2Cu_3O_{6+x}$ (R=Tm, Lu) in the vicinity of the Néel temperature $T_N$ [18]; it was found that the *c*-axis transport occurs through two conduction channels, one of which is essentially blocked by the AF ordering. As a result of such blocking, a steep increase in both $\rho_c$ and the anisotropy ratio $\rho_c/\rho_{ab}$ was observed upon cooling below $T_N$. In a study of the *ab*-plane and *c*-axis magnetoresistance (MR) of a series of heavily underdoped $YBa_2Cu_3O_{6+x}$ (Y-123) crystals, we found that the *c*-axis MR undergoes a drastic change in the vicinity of the Néel temperature [2]. Also, quite unexpectedly, no feature associated with the AF ordering was found in the transverse *ab*-plane MR [*I* // *a(b)*; *H* // *c*]. These new data indicate that the development of the AF correlations and the formation of the long-range Néel order have a profound influence only on the charge transport across the $CuO_2$ planes, leaving the in-plane transport unchanged. Moreover, the data show that the longitudinal *c*-axis MR (*H* // *c*) is apparently governed by the AF fluctuations even in the temperature range *above* $T_N$, indicating that the spin fluctuations are playing a major role in the *c*-axis transport regardless of the presence of the Néel order.

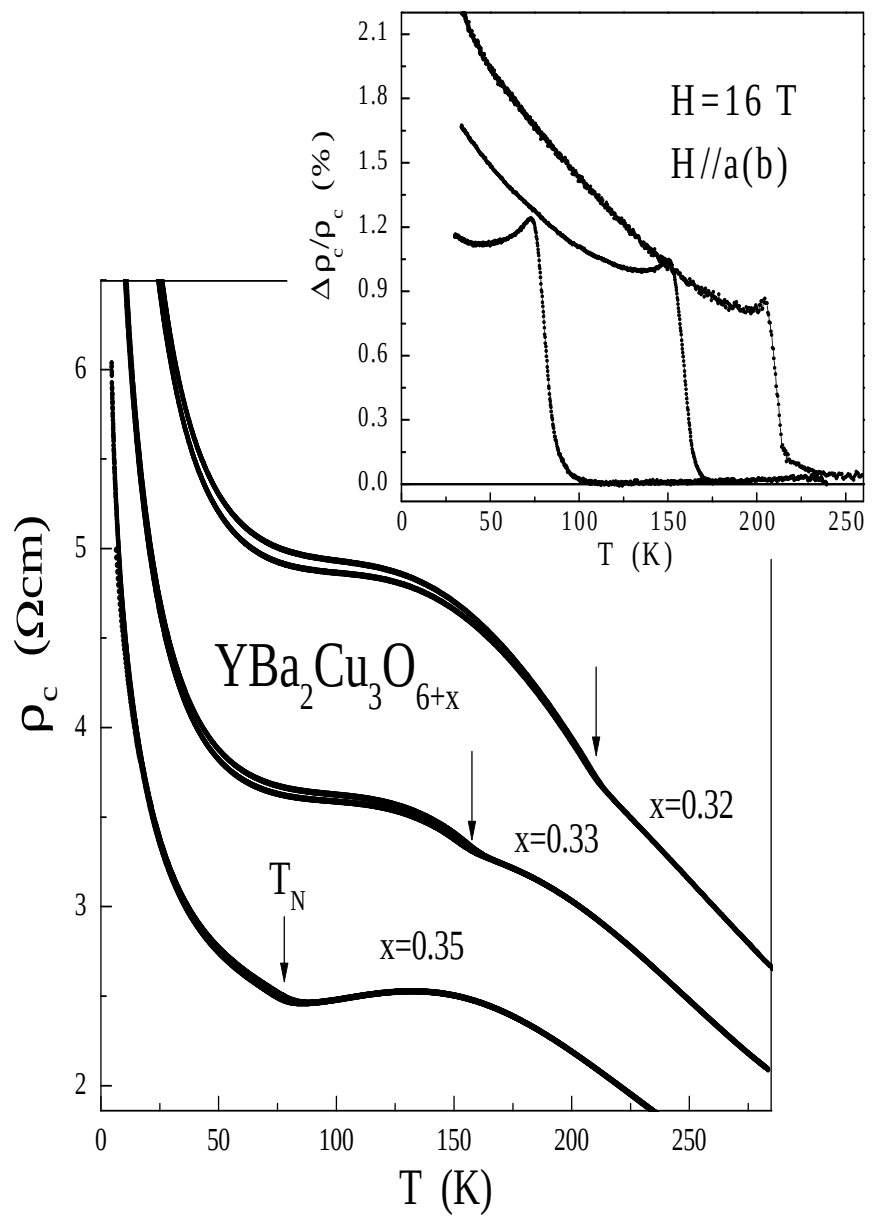

Fig. 2. $\rho_c$ *(T)* of an Y-123 single crystal at three different oxygen contents. Curves for *H*=0 and 16 T [*H* // *a(b)*] are shown. The kink on the resistivity curves marks the AF transition. Inset: *T* dependences of the *c*-axis transverse MR [*H* // *a(b)*] at 16 T .

The MR measurements are performed either by sweeping temperature (controlled by a Cernox resistance sensor) under constant magnetic fields up to 16 T, or by sweeping the field at a fixed temperature stabilized by a capacitance sensor to an accuracy of about 1 mK. The latter method allows measurements of $\Delta\rho/\rho$ as small as $10^{-5}$ at 10 T. In Fig. 2 we present a set of $\rho_c$ curves obtained for the same Y-123 single crystal at slightly different oxygen contents in the AF region. The rise in $\rho_c$ induced by the AF transition becomes more and more evident as $T_N$ is lowered. Note that both the zero-field and the 16 T data are shown in the main panel of Fig. 2, where the difference between the two becomes noticeable below $T_N$. The inset of Fig. 2 demonstrates an unusual behavior of the *c*-axis MR, in which a step-like increase in $\Delta\rho_c/\rho_c$ is observed upon cooling through $T_N$. Except for a small difference in the MR step width, which is obviously related to the width of the AF transition, this striking feature is very reproducible within a set of Y-123 crystals. One would expect that the Néel transition in Y-123, like other phase transitions associated with the magnetic subsystem, is considerably affected by the application of magnetic fields. If a strong magnetic field suppresses AF order and lowers $T_N$, the *c*-axis MR should become negative, because $\rho_c$ is enhanced below $T_N$; however, we have found that the *c*-axis MR is *positive*, which is opposite to such a naive expectation. Also surprisingly, when we look at the in-plane transport, we do not find any anomaly that can be associated with the Néel transition, and the *ab*-plane MR, $\Delta\rho_{ab}/\rho_{ab}$, is always smooth in the vicinity of $T_N$.

One would naturally ask whether the out-of-plane transport is sensitive exclusively to the long-range order arising below $T_N$; if the short-range AF correlations above $T_N$ also contribute to $\rho_c$ and its MR, we may expect that the AF fluctuations play an essential role not only in the AF compositions but also in the superconducting compositions. To clarify this point, we obtained more precise MR data with the field-sweeping measurements to investigate the behavior above $T_N$, where the MR becomes very small. The *T* dependences of $\gamma_\perp H^2$ and $\gamma_{//} H^2$ components of $\Delta\rho_c/\rho_c$ (for $H \perp c$ and *H* // *c*, respectively) presented in Fig. 3 depict the qualitative difference in the MR behavior for the two directions of the magnetic field. For *H* // *ab* [Fig. 3(a)], the *c*-axis MR changes at $T_N$ in a step-like manner by up to two orders of magnitude. The step separates regions below and above $T_N$ with relatively weak dependence of the MR on temperature. This behavior implies that the sensitivity to the magnetic field appears abruptly with the onset of the long-range AF order. On the other hand, for *H* // *c* [Fig. 3(b)] we observe a MR peak at $T_N$, which is accompanied by a tail spreading to far above $T_N$. The MR as a function of temperature has no discontinuity at $T_N$ and one can infer from Fig. 3(b) that the *c*-axis MR grows as $T^{-k}$ with decreasing temperature until the Néel transition interrupts this tendency. However, the right-hand side of the MR peak for *H* // *c* (the $T^{-k}$ behavior) apparently shifts with $T_N$ when the *x* is changed, which indicates its relation to the AF ordering. Therefore, we can conclude that a mechanism associated obviously with the AF fluctuations dominates the *c*-axis MR in a wide temperature range above $T_N$ as well. This observation clearly demonstrates that the short-range AF correlations play an essential role in the out-of-plane transport.

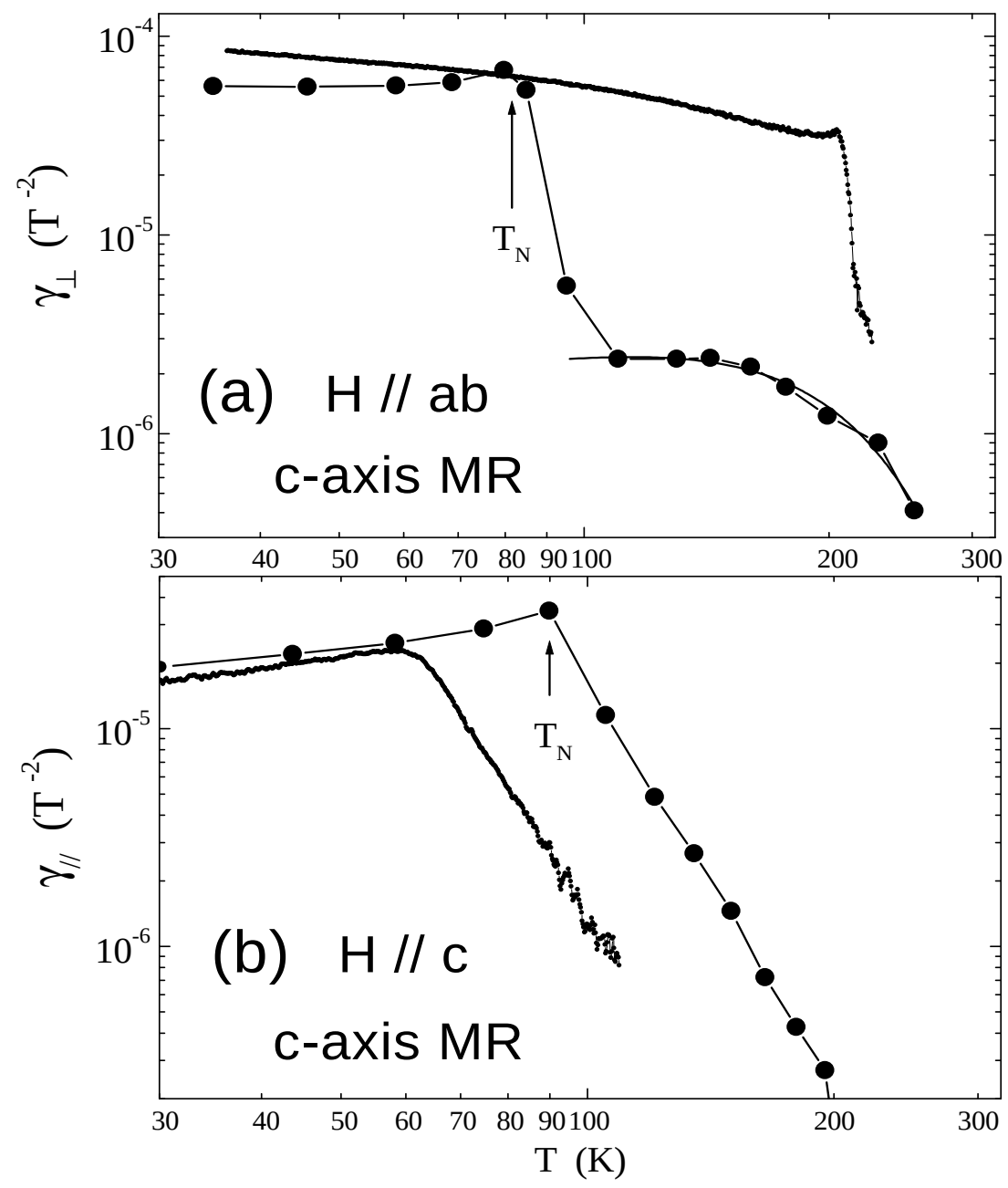

Fig. 3. *T* dependences of the $\gamma_\perp H^2$ term in the *c*-axis MR with (a) *H* // *ab* and (b) *H* // *c*. A Lu-123 crystal was used for the field sweep measurements (solid circles). The MR data for Y-123 under sweeping temperature are shown for comparison (dots).

The contrasting behavior of the *ab*- and *c*-axis MR in the heavily underdoped Y-123 indicates that changes that occur in the spin subsystem at $T_N$ are influential only on the electron transport across the $CuO_2$ planes and apparently not on the in-plane one. It is known that the heavily underdoped Y-123 above $T_N$ possesses well-developed dynamic AF correlations in the $CuO_2$ planes [19] and the Néel temperature actually corresponds to the establishment of the AF order along the *c*-axis. It is possible that the freezing of the spin degrees of freedom below $T_N$ causes an increase in $\rho_c$, if the spin fluctuations assist the electron hopping between the $CuO_2$ planes. Since an increase in $\rho_c$ also takes place when the magnetic field is applied, one can infer that the field suppression of the spin fluctuations is likely to be the main source of the positive *c*-axis MR in our heavily underdoped Y-123. Also, the dramatic changes in the out-of-plane transport associated with the evolution of the magnetic state might suggest that it is the spin subsystem that is responsible for the charge confinement within the $CuO_2$ planes.

---

* Work done in collaboration with A. N. Lavrov, Kouji Segawa, J. Takeya, G. S. Boebinger, A. Passner, N. L. Wang, C. Geibel, and F. Steglich.